\documentclass{ws-ijmpa}
\usepackage[super,compress]{cite}
\usepackage{graphicx}
\begin{document}
\markboth{Koichi Hirano}
{Neutrino masses from CMB B-mode polarization and cosmic growth rate}

%
\catchline{}{}{}{}{}
%

\title{
Neutrino masses from CMB B-mode polarization and \\
cosmic growth rate
}

\author{KOICHI HIRANO
}

\address{
Department of Primary Education, Tsuru University,\\
3-8-1, Tahara, Tsuru, Yamanashi 402-8555, Japan
\\
k\_hirano@tsuru.ac.jp}



\maketitle


\begin{abstract}
Constraints on neutrino masses are estimated based on future observations of the cosmic microwave background (CMB), which includes the B-mode polarization produced by CMB lensing from the Planck satellite, and the growth rate of cosmic structure from the Euclid redshift survey by using the Markov-Chain Monte-Carlo (MCMC) method. The error in the bound on the total neutrino mass is estimated to be $\Delta\sum m_{\nu} = 0.075$ eV with a 68\% confidence level. By using the growth rate rather than the galaxy power spectrum, accurate constraints are obtained, since the growth rate is less influenced by the uncertainty regarding galaxy bias than by the galaxy power spectrum.

\end{abstract}

\ccode{PACS numbers: 98.80.-k, 14.60.Pq, 98.70.Vc, 98.65.Dx}


\section{Introduction}

The standard model of particle physics assumes that neutrinos are massless. However, neutrino oscillation experiments indicate that neutrinos have nonzero masses. Experimental mass differences between the neutrinos are $|\Delta m_{21}^2| = 7.59^{+0.19}_{-0.21}\times 10^{-5}{\rm eV}^2$ \cite{aharmim2008} and $|\Delta m_{32}^2| = 2.43^{+0.13}_{-0.13}\times 10^{-3}{\rm eV}^2$ \cite{adamson2008}. However, the absolute masses and hierarchical structure have not yet been determined. Determinations of these are essential to advance physics beyond the standard model.

Terrestrial experiments such as those regarding tritium beta decay \cite{thummler2011} and neutrinoless double-beta decay \cite{Gomez-Cadenas2011} give upper bounds on the absolute neutrino masses. Cosmological observations could further constrain neutrino properties by providing a more stringent bound on the total neutrino mass $\sum m_{\nu}$ and the effective number of neutrino species $N_\nu$.

Cosmic microwave background (CMB) anisotropies are mainly generated until the eras before the last scattering surface of the decoupling epoch (redshift $z \sim 1089$). Therefore, if neutrinos are as massive as $\sum m_{\nu} \hspace{0.3em}\raisebox{0.4ex}{$>$}\hspace{-0.75em}\raisebox{-.7ex}{$\sim$}\hspace{0.3em} 1.5$ eV, they become nonrelativistic before the recombination epoch. In such a case, a finite-mass neutrino will significantly affect the CMB spectrum. For masses below $\sum m_{\nu} \hspace{0.3em}\raisebox{0.4ex}{$<$}\hspace{-0.75em}\raisebox{-.7ex}{$\sim$}\hspace{0.3em} 1.5$ eV, the neutrinos alter the CMB spectrum primarily through their effect on the angular diameter distance to the last scattering surface. In this case, the effect is degenerate with other cosmological parameters, such as the matter energy density parameter $\Omega_{m}$ and the Hubble constant $h$ \cite{ichikawa2005}. Other cosmological probes complementary to the CMB are needed to break the parameter degeneracy in order to study the small mass scales of neutrinos.

The B-mode polarization due to CMB lensing provides detailed information. This polarization is very sensitive to neutrino masses smaller than $0.1$ eV. This resolution is indispensable when distinguishing between a normal and an inverted hierarchy. The gravitational lensing B-modes have recently been detected for the first time by a study that used data from the SPTpol detector of the South Pole Telescope \cite{hanson2013}. The first detection for the primordial B-mode polarization of the cosmic microwave background (CMB) was reported by astronomers working on the Background Imaging of Cosmic Extragalactic Polarization (BICEP2) telescope at the South Pole \cite{ade2014}. Validity of the results of the BICEP2 is being checked regarding foreground radiation. The angular power spectrum of polarized dust emission by the Planck has been released \cite{adam2014}, but detailed polarization data of CMB from the Planck are scheduled to be reported in December 2014.

The main effects of massive neutrinos on the growth of matter density perturbations arise from two physical mechanisms \cite{lesgourgues2006a}. In the first mechanism, a massive neutrino becomes nonrelativistic at the transition temperature, and contributes to the energy density of cold dark matter. This changes the matter-radiation equality time and the expansion rate of the universe. In the second mechanism, the matter power spectrum is suppressed at small scales by neutrino free-streaming. Neutrinos travel at the speed of light as long as they are relativistic, and the free-streaming scale is nearly equal to the Hubble horizon. Therefore, the free-streaming effect suppresses perturbations below such scales.

Neutrino masses from cosmology have been studied by combining observations of CMB anisotropies with galaxy clustering \cite{saito2008,saito2009,saito2011}, weak lensing \cite{ichiki2009}, and the Lyman-$\alpha$ Forest \cite{viel2006,seljak2006}. The Planck CMB temperature power spectrum with WMAP polarization constrains the sum of the neutrino masses to $\sum m_{\nu}<$ 0.933 eV (95\% CL) \cite{planck2013_16}. By combining the Planck temperature data with WMAP polarization, the high-resolution CMB data, and the distance measurements from the baryon acoustic oscillations (BAO), a robust upper bound of $\sum m_{\nu} <$ 0.230 eV has been reported \cite{planck2013_16}.
By focusing on the ongoing and future observations of both the 21-cm line and the CMB B-mode polarization, the sensitivities to the effective number of neutrino species, total neutrino mass, and neutrino mass hierarchy have been studied \cite{oyama2013}.

In this paper, Planck data \cite{planck2013_1,planck2013_15} from ongoing CMB observations including the B-mode polarization from CMB lensing are used, and the Euclid mission \cite{amendola2012} is adopted for the future observations of the growth rate of cosmic structure from the redshift survey. Robust constraints are estimated by using the growth rate rather than the galaxy power spectrum, since many uncertainties regarding galaxy bias remain when the galaxy power spectrum is calculated from the matter power spectrum. The errors in the bounds on the total neutrino mass $\Delta\sum m_{\nu}$ are accurately estimated by comparing the observational data with the models.

This paper is organized as follows. In the next section, the model used here is summarized. In Section \ref{sec_observ}, the Planck satellite and the Euclid mission, and mock observational data used in this study are described. In Section \ref{sec_analysis}, our likelihood analysis by using the Markov-Chain Monte-Carlo (MCMC) method with the mock data is shown. Finally, results are given in Section \ref{sec_results}.

\section{Model}

Here a flat $\Lambda$CDM model with two additional parameters of the total neutrino mass $\sum m_{\nu}$ and the effective number of neutrino species $N_\nu$ is used. The neutrino mass is related to the neutrino density parameter as $\Omega_{\nu}h^2 = \Sigma m_\nu/(93.04~{\rm eV})$. The CMB temperature is taken to be $T_{\rm CMB} = 2.7255~{\rm K}$ \cite{fixsen2009}. The primordial helium fraction $Y_{\rm P}$ is a function of $\Omega_{\rm b}h^2$ and $N_{\nu}$ and uses the big bang nucleosynthesis consistency condition \cite{ichikawa2006,hamann2008}.

\section{Observational data} \label{sec_observ}

Planck is the third CMB observation satellite, following COBE and WMAP. It is possible to take all of the information in the CMB temperature anisotropies to measure the polarization of the CMB anisotropies with high accuracy. Planck provides the thermal history of the universe during the period of the formation of the first stars and galaxies. It is possible to detect the signature of gravitational waves generated during inflation by polarization measurements \cite{planck2006}.

Data from Planck are used for the CMB observations that included B-mode polarization caused by CMB lensing. The satellite was launched in May 2009. In March 2013, initial cosmology results based on the first 15.5 months of operation were released with an analysis of the temperature data \cite{planck2013_1,planck2013_15,planck2013_16,planck2013_17}. Detailed polarization data are scheduled to be released in December 2014. In the current study, mock data of the polarization of the CMB anisotropies generated by the FuturCMB code \cite{perotto2006} are used. Experimental specifications assumed in the computation are summarized in Table \ref{planck}. The maximum multipoles ($l_{max}$ = 2500) for Planck are used.

\begin{table}[h!]
\tbl{Experimental specifications of the CMB projects. Here $f_{\rm sky}$ is the observed fraction of the sky, $\nu$ is the observation frequency, $\theta_{\rm FWHM}$ is the angular resolution defined as the full width at half maximum, $\Delta_{\rm T}$ is the temperature sensitivity per pixel, and $\Delta_{\rm P}$ is the polarization sensitivity per pixel.}
{\begin{tabular}{c c c c c c}
\hline
\hline
Experimental Parameters & $f_{\rm sky}$ & $\nu$ & $\theta_{\rm FWHM}$ & $\Delta_{\rm T}$ & $\Delta_{\rm P}$ \\
 & & $[{\rm GHz}]$ & $[']$ & $[\mu{\rm K}]$ & $[\mu{\rm K}]$ \\
\hline
Data from Planck \cite{planck2013_1} & 0.73 & 100  & 9.66  & 6.8  & 10.9 \\ 
       &      & 143  & 7.27  &  6.0   & 11.4 \\
       &      & 217  & 5.01  &  13.1  & 26.7 \\
\hline
\hline
\end{tabular} \label{planck}}
\end{table}

Euclid \cite{amendola2012} is a European Space Agency medium class mission that is scheduled to be launched in 2019. The main purpose of Euclid is to determine the origin of the accelerated expansion of the universe. Euclid will research the expansion history and the evolution of cosmic structures by measuring redshifts of galaxies and the distribution of clusters of galaxies over a large portion of the sky. Although its main subject of research is the nature of dark energy, Euclid will cover topics such as cosmology, galaxy evolution, and planetary research.

\begin{figure}[h!]
\centerline{\includegraphics[width=\textwidth]{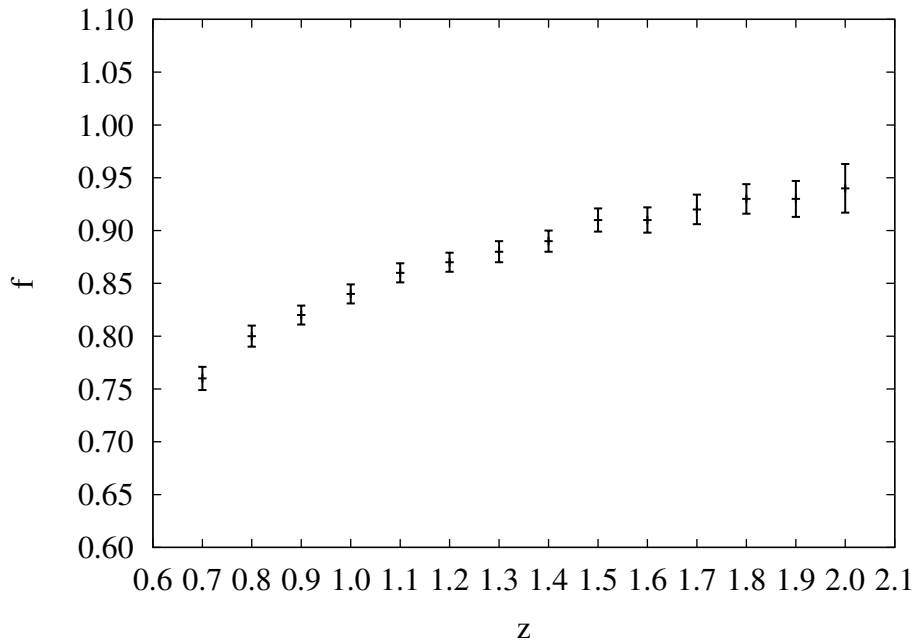}}
\caption{Plot of the mock data of the cosmic growth rate.
\label{fig1}}
\end{figure}

In this study, Euclid parameters are adopted for the growth rate observations. The growth rate can be parameterized by the growth index $\gamma$, as defined by $f = {\Omega_m}^\gamma$. Mock data of the structural growth rate are created in accordance with the $1\sigma$ marginalized errors of the growth rate that will be used by Euclid, which are shown in Table 4 in the study by Amendola et al. \cite{amendola2012}. Table \ref{euclid} lists the $1\sigma$ marginalized errors for the growth rates in each redshift bin based on Table 4 in the study by Amendola et al. \cite{amendola2012}. In Fig. \ref{fig1}, the mock data of the cosmic growth rate used in this current study are plotted.

\begin{table}[h!]
\tbl{$1\sigma$ marginalized errors for the growth rates in each redshift bin based on Table 4 in the study by Amendola et al. \cite{amendola2012}. Here $z$ represents the redshift and $\sigma_{f_g}$ represents the $1\sigma$ marginalized errors for the growth rates.}
{\begin{tabular}{c c c c c c}
\hline
\hline
Experimental Parameters & $z$ & $\sigma_{f_g}$(ref.) \\
\hline
Data from Euclid \cite{amendola2012} & 0.7 & 0.011 \\
 & 0.8 & 0.010 \\
 & 0.9 & 0.009 \\
 & 1.0 & 0.009 \\
 & 1.1 & 0.009 \\
 & 1.2 & 0.009 \\
 & 1.3 & 0.010 \\
 & 1.4 & 0.010 \\
 & 1.5 & 0.011 \\
 & 1.6 & 0.012 \\
 & 1.7 & 0.014 \\
 & 1.8 & 0.014 \\
 & 1.9 & 0.017 \\
 & 2.0 & 0.023 \\
\hline
\hline
\end{tabular} \label{euclid}}
\end{table}

The mock data are used to calculate the statistical $\chi^2$ function. $\chi^2$ for the growth rate is defined as
\begin{equation}
\displaystyle \chi_f^2 = \sum_{i=1}^{14}\frac{(f_{theory}(z_i)-f_{obs}(z_i))^2}{\sigma_{f_g}(z_i)^2}
\end{equation}

where $f_{obs}(z_i)$ are the future observational (mock) data of the growth rate. The theoretical growth rate $f_{theory}(z_i)$ is computed as
\begin{equation}
 f = \frac{d\ln{\delta}}{d\ln{a}}
\end{equation}
where $\delta$ is the matter density fluctuations and $a$ is the scale factor.

The estimated errors from observational technology are known, but the center value of the future observations is not known. Therefore, in this analysis, the error in the bound on the total neutrino mass $\Delta\sum m_{\nu}$ is significant, but the most likely value of the total neutrino mass $\sum m_{\nu}$ is not.

\section{Likelihood analysis} \label{sec_analysis}

Using the above data, the Markov-Chain Monte-Carlo (MCMC) method \cite{lewis2002,cosmomc} is used to search the cosmological parameter estimations in the multidimensional parameter space of cosmological observables. The error bounds on the cosmological parameters are estimated. Recently, the Fisher matrix has become a standard to estimate errors in cosmological parameters for future observations. However, when the phenomena are not Gaussian distributed (such as in the case of strong parameter degeneracies), the Fisher matrix formalism loses validity, as described by Perotto and colleagues \cite{perotto2006}. Because all parameter likelihoods cannot always be approximated with a Gaussian distribution, Monte Carlo simulations based on the publicly available CosmoMC code \cite{lewis2002,cosmomc} are used with the mock observational data. The cosmological parameter ranges that are explored with the MCMC method are listed in Table \ref{space}.

\begin{table}[h!]
\tbl{The prior ranges that are explored. Here $\Omega_{\rm b}h^2$ is the current baryon density, $\Omega_{\rm c}h^2$ is the current cold dark matter density, $100\theta_{\rm MC}$ is 100 $\times$ the approximation to $r_*/D_{\rm A}$ (CosmoMC), $\tau$ is the Thomson scattering optical depth due to reionization, $n_{\rm s}$ is the scalar spectrum power-law index, $\ln{(10^{10}A_{\rm s})}$ is the log power of the primordial curvature perturbations, $f_\nu$ is the fraction of the dark matter that is in the form of massive neutrinos, and $N_{\nu}$ is the effective number of neutrino-like relativistic degrees of freedom.}
{\begin{tabular}{l c}
\hline
\hline
Parameter~ & Prior Range \\
\hline
$\Omega_{\rm b}h^2$ & 0.005 $\rightarrow$ 0.1 \\
$\Omega_{\rm c}h^2$ & 0.01 $\rightarrow$ 0.99 \\
$100\theta_{\rm MC}$ & 0.3 $\rightarrow$ 10 \\
$\tau$ & 0.01 $\rightarrow$ 0.8 \\
$n_{\rm s}$ & 0.5 $\rightarrow$ 1.5 \\
$\ln{(10^{10}A_{\rm s})}$ & 0.5 $\rightarrow$ 6.0 \\
$f_\nu$ & 0 $\rightarrow$ 1.0 \\
$N_{\nu}$ & 0.1 $\rightarrow$ 8.0 \\
\hline
\hline
\end{tabular} \label{space}}
\end{table}

``High accuracy default'' and ``accuracy level'' 3 are implemented in CAMB \cite{lewis2000,camb}. The chains have 1,000,000 points in CosmoMC.

\section{Results} \label{sec_results}

\begin{figure}[h!]
\centerline{\includegraphics[width=\textwidth]{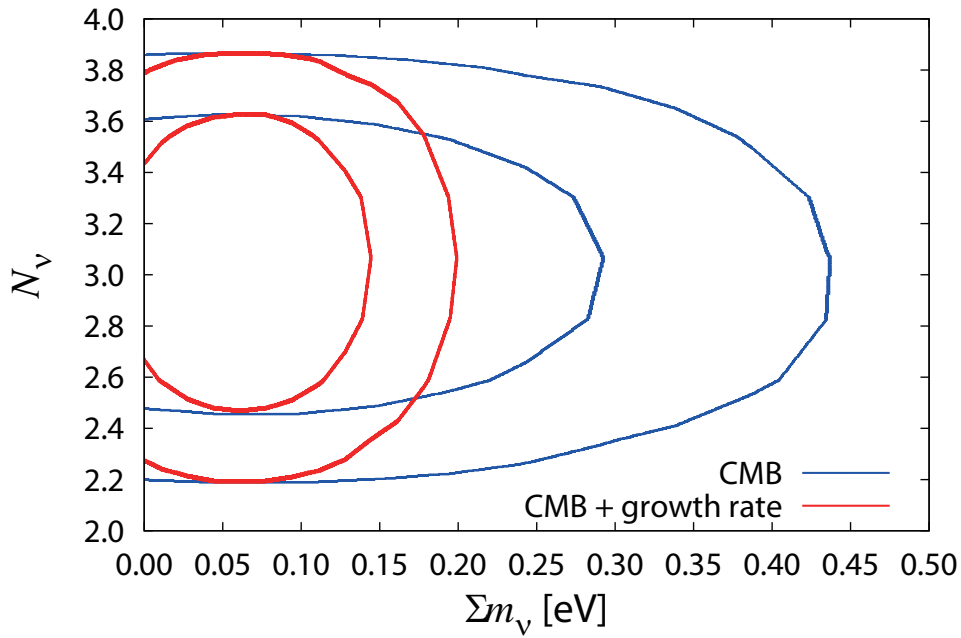}}
\caption{Probability contours in the ($\sum m_{\nu}$, $N_{\nu}$)-plane. The contours show the 1$\sigma$ (68\%) and 2$\sigma$ (95\%) confidence limits, for the observational data of the CMB with the B-mode polarization (blue line) and CMB including B-mode plus the growth rate (red line).
\label{fig2}}
\end{figure}

In Fig. \ref{fig2}, the probability contours in the ($\sum m_{\nu}$, $N_{\nu}$)-plane are plotted. The fiducial value of the total neutrino mass is $\sum m_{\nu} = 0.06$ eV and the effective number of neutrino-like relativistic degrees of freedom is $N_{\nu} = 3.046$, whereas the other parameters are marginalized. Bule lines are constraints from the observational data of the CMB with the B-mode polarization (Planck), and red lines are constraints from the observational data of the CMB with the B-mode polarization (Planck) and that with the addition of the growth rate (Euclid). The contours show the 1$\sigma$ (68\%) and 2$\sigma$ (95\%) confidence limits.

The following error on the bounds of the total neutrino mass was obtained after using the data of the CMB with the B-mode polarization (Planck):
\begin{equation}
\Delta\sum m_{\nu} = 0.20~{\rm eV~~(68\% CL)~~~(CMB)}
\end{equation}
Furthermore, the following more stringent constraint was obtained after using the data of the CMB with the B-mode polarization (Planck) and growth rate (Euclid):
\begin{equation}
\Delta\sum m_{\nu} = 0.075~{\rm eV~~(68\% CL)~~~(CMB + growth~rate)}
\end{equation}

It was found that the addition of the growth rate data had little effect on the constraints of the effective number of neutrino-like relativistic degrees of freedom $N_{\nu}$ as shown in Fig. \ref{fig2}.  

\begin{figure}[h!]
\centerline{\includegraphics[width=\textwidth]{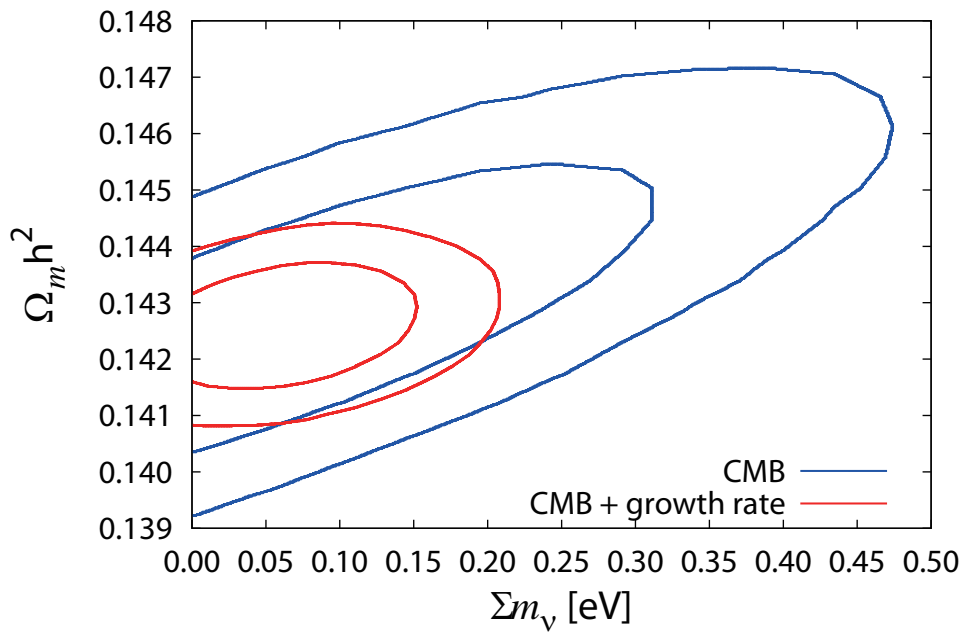}}
\caption{Probability contours in the ($\sum m_{\nu}$, $\Omega_{m}h^2$)-plane. The contours show the 1$\sigma$ (68\%) and 2$\sigma$ (95\%) confidence limits, for the observational data of the CMB with the B-mode polarization (blue line) and CMB including B-mode plus the growth rate (red line).
\label{fig3}}
\end{figure}

In Fig. \ref{fig3}, the probability contours in the ($\sum m_{\nu}$, $\Omega_{m}h^2$)-plane are plotted. The fiducial value of the total neutrino mass is $\sum m_{\nu} = 0.06$ eV and the current energy density of matter is $\Omega_{m}h^2 = 0.1426$, whereas $h$ is the Hubble parameter. The other parameters are marginalized.

It was found that the parameter degeneracies of $\sum m_{\nu}$ and $\Omega_{m}h^2$ were efficiently broken after adding the CMB with the B-mode polarization (Planck) data to the growth rate (Euclid) data because the energy density parameter of matter $\Omega_{m}h^2$ was stringently constrained by the growth rate data as shown in Fig. \ref{fig3}.

In summary, by combining the ongoing CMB observations, which include the B-mode polarization caused by CMB lensing (Planck), and the future observations of the growth rate of cosmic structures (Euclid), the error in the bound of the total neutrino mass was estimated to be $\Delta\sum m_{\nu} = 0.075$ eV with a 68\% confidence level. This result of the error in the bound of the total neutrino mass is nearly independent of fiducial value because the error in the bound is estimated based on the experimental specifications of the Planck (Table \ref{planck}) and the Euclid (Table \ref{euclid}).

The Markov-Chain Monte-Carlo (MCMC) method was used to estimate the error bounds on the cosmological parameters, because the Fisher matrix formalism loses validity when the phenomena are not Gaussian distributed.
The growth rate was used rather than the galaxy power spectrum, since the growth rate is less influenced by the uncertainty regarding galaxy bias than by the galaxy power spectrum. As a result, more ``accurate'' and ``conservative'' constraints were obtained than the results in other similar papers \cite{hamann2012,carbone2012,joudaki2012,santos2013,takeuchi2014,font-ribera2014}. The ``width'' of the probability distribution was less than $75$ meV after using future observations of the B-mode polarization and cosmic growth rate.

It is known that the total neutrino mass is $\sum m_{\nu} \hspace{0.3em}\raisebox{0.4ex}{$>$}\hspace{-0.75em}\raisebox{-.7ex}{$\sim$}\hspace{0.3em} 0.1$ eV in case of an inverted hierarchy. Hence, depending on the fiducial value of the total neutrino mass $\sum m_{\nu}$, it is possible that an inverted hierarchy is rejected.

\section*{Acknowledgments}

This work was supported by a Grant-in-Aid for Scientific Research from the Japan Society for the Promotion of Science (Grant Number 25400264 (KH)).


\bibliography{koichi}

\end{document}